\documentstyle[aps,prl,psfig,epsfig,multicol]{revtex}

\begin{document}

\title{Anisotropic Magnetoconductance in Quench-Condensed Ultrathin Beryllium Films }

\draft


\author{E. Bielejec, J. Ruan, and Wenhao Wu}
\address{Department of Physics and Astronomy, University of Rochester,}
\address{Rochester, New York 14627}

\date{\today}

\maketitle

\begin{abstract}

Near the superconductor-insulator (S-I) transition,
quench-condensed ultrathin Be films show a large
magnetoconductance which is highly anisotropic in the direction of
the applied field. Film conductance can drop as much as seven
orders of magnitude in a weak perpendicular field ($\le$ 1 T), but
is insensitive to a parallel field in the same field range. We
believe that this negative magnetoconductance is due to the field
de-phasing of the superconducting pair wavefunction. This idea
enables us to extract the finite superconducting phase coherence
length, L$_{\phi}$, in nearly superconducting films. Our data
indicates that this local phase coherence persists even in highly
insulating films in the vicinity of the S-I transition.

\end{abstract}

\pacs{PACS numbers: 73.50.-h, 71.30.+h, 74.50.+r, 74.80.Bj}

\begin{multicols}{2}

The superconductor-insulator (S-I) transition remains a
controversial subject after nearly two decades of intense
research. Experimental systems have generally been grouped into
two seemingly different categories, displaying somewhat different
features. These are granular versus uniform films. In apparently
granular films \cite{Ekinci}, the transition temperature, T$_{c}$,
and the energy gap, $\Delta$, for the individual grains are
essentially constant throughout the S-I transition \cite{White}.
It is well established that superconductivity is destroyed by the
breakdown of long-range phase coherence between the grains
\cite{White,Orr}. Granular films tend to lose their zero
resistance state at a normal state sheet resistance, R$_{N}$, that
is close to the quantum pair resistance, R$_{Q}$ = h/(2e)$^{2}$
$\approx$ 6.5 k$\Omega$. In addition, granular films display
quasireentrance in films not far from being superconducting: film
resistance initially drops as the temperature is cooled below
T$_{c}$, but eventually increases at low temperatures. A truly
insulating state is not seen in granular films until R$_{N}$ $\gg$
R$_{Q}$. Films considered to be uniformly disordered, such as
quench-condensed Bi/Ge and Pb/Ge films \cite{Haviland}, amorphous
InO$_{x}$ films \cite{Hebard}, and $\em{a}$-MoGe films
\cite{Yazdani}, undergo a much sharper S-I transition. On the
superconducting side, T$_{c}$ decreases with decreasing film
thickness, approaching zero at the S-I transition. Tunneling
experiments have suggested \cite{Valles} that in Bi/Ge and Pb/Ge
films the superconducting gap, $\Delta$, decreases with decreasing
film thickness until the pair wavefunction,
$\Delta^{1/2}$e$^{i\phi}$, with $\phi$ being the phase of the
order parameter, vanishes and the film becomes insulating. These
results suggest that the S-I transition is driven by the vanishing
of the superconducting gap. In the alternative "dirty-boson" model
\cite{Fisher}, superconductivity is suppressed by $\em{phase}$
fluctuations, and the Cooper pairs persist even on the insulating
side of the transition. In recent years, the "dirty-boson" model
has been applied nearly exclusively to explain the scaling
analyses of film resistance in the disorder-driven and
field-driven S-I transitions in uniform films
\cite{Hebard,Yazdani,Liu,Markovic}, although the existence of
Cooper pairs in the insulating states of these films has yet to be
demonstrated. The "dirty-boson" model is, in fact, expected to
describe best the S-I transition in granular films, since the
amplitude of the order parameter is well defined on both side of
the S-I transition in granular films.

The key concept of the "dirty-boson" model is that phase
fluctuations drive a continuous S-I transition. Thus, in the
vicinity of the S-I transition, there should exist a finite
superconducting phase coherence length, L$_{\phi}$, even on the
insulating side of the transition. This L$_{\phi}$ should scale
with the correlation length of the transition and should diverge
approaching the transition. In this Letter, we report on
magnetoconductance (MC) measurements in the vicinity of the S-I
transition in quench-condensed ultrathin Be films. As we argue
below, this MC study provides the first direct measurement of
L$_{\phi}$ in insulating films. We have observed that, for a given
insulating film, L$_{\phi}$ drops as temperature is lowered. This
underscores the competition between localization and
superconductivity. Approaching the superconducting state with
increasing film thickness, we have found L$_{\phi}$ to grow
drastically.

Our ultrathin Be films were quench-condensed onto bare glass
substrates which were held near 20 K during the evaporations. We
chose Be mainly for two reasons. First, earlier studies have
suggested \cite{Yatsuk} that quench-condensed Be films are nearly
amorphous. Scanning force microscopy studies of our Be films,
after warming up to room temperature, have found no observable
granular structure down to 1 nm. This indicates that the length
scale of disorder in these Be films must be much smaller than the
typical grain size in apparently granular films. Second, Be has a
very weak spin-orbit coupling \cite{Tedrow}. As a result, a
magnetic field applied parallel to the film plane,
H$_{\parallel}$, couples to electron spin only and it does not
couple to the orbital motion of the electrons. However, a
perpendicular field, H$_{\perp}$, couples to both. Thus the MC
\begin{figure}
\centerline{\epsfig{file=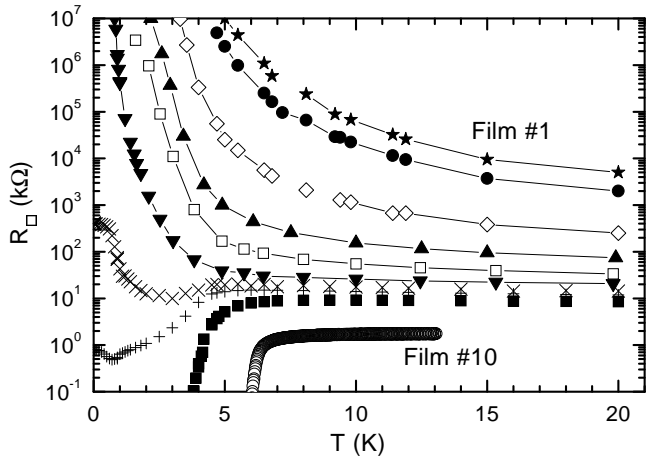,width=8.5cm}}
\caption{Selected curves of R$_{\Box}$ versus temperature measured
on one film area following a series of deposition steps to
increase film thickness. Curves from top to bottom are labeled as
Film $\#$1 to Film $\#$10, respectively. The thickness for these
films changed from 4.6 {\AA} to 15.5 \AA.} \label{Figure 1}
\end{figure}
\noindent can be highly anisotropic in the direction of the
applied field. In our Be films, the MC is negative and varies as
much as seven orders of magnitude in weak H$_{\perp}$ up to 1 T,
but it is insensitive to H$_{\parallel}$ in the same field regime.
The low-field MC in our films is thus clearly an orbital effect.
We, therefore, believe that this negative and highly aniostropic
MC provides a direct measurement of the superconducting phase
coherence length, L$_{\phi}$, as was suggested by Barber and Dynes
\cite{Barber} in a MC study of superconducting granular Pb films.
This method can eventually be used to measure the divergence of
L$_{\phi}$ as films cross the S-I transition with varying film
thickness.

It should be pointed out that the length scale of the disorder in
films considered uniformly disordered is still not understood.
Even if microscopy techniques fail to reveal any granular
structure, there still can exist metallic clusters, which can
support superconductivity and which are connected electrically by
relatively narrow and insulating or metallic links. For example,
the Ge underlayer in Bi/Ge and Pb/Ge films may produce tunneling
channels connecting the superconducting clusters. Recently,
Kapitulnik and collaborators \cite{Yazdani} have proposed that,
near the S-I transition in $\em{a}$-MoGe, there exist both
insulating and superconducting puddles, with transport being
dominated by tunneling or hopping between them. Presumably, the
typical size of the superconducting puddles grows approaching the
S-I transition and eventually become the longest length scale of
the system. Another example is the InO$_{x}$ films studied by
Hebard $\em{et}$ $\em{al.}$ \cite{Hebard}, which are believed to
be amorphous, yet they display the quasireentrant behavior of
granular films. Thus it is not clear as to what are the
fundamental differences between the S-I transitions observed in
uniform and granular films, other than that the different
morphologies may lead to different universality classes of the
transitions.

The details regarding our quench-condensation appa-
\begin{figure}
\centerline{\epsfig{file=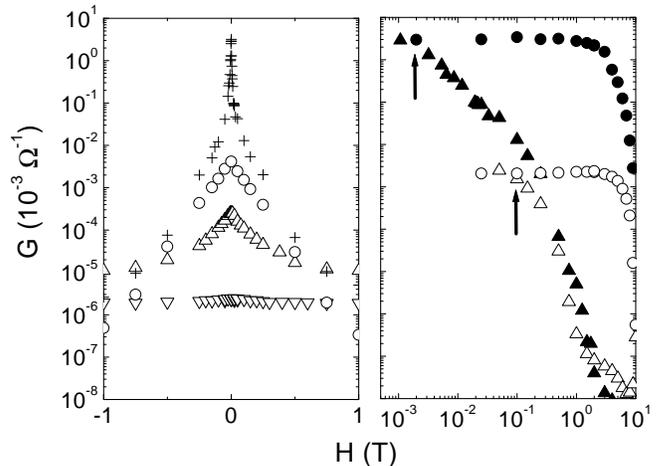,width=8.5cm}} \caption{(a):
MC measured in H$_{\perp}$. Up-triangles and down-triangles are
measured on Film $\#$6 at 1.5 K and 1.0 K, respectively. MC was
not measured on this film below 1.0 K because the film was too
resistive. Circles are from Film $\#$7 at 100 mK. Crosses are from
Film $\#$8 at 100 mK. (b): Comparison of MC data in
H$_{\parallel}$ (filled and open circles) and H$_{\perp}$ (filled
and open triangles) at 100 mK on a log-log scale. Filled symbols
are for Film $\#$8. Open symbols are for Film $\#$7. The data in
H$_{\perp}$ are copied from (a). The arrows indicate the crossover
fields, H$^{\ast}$, at which film conductance starts dropping with
increasing field.} \label{Figure 2}
\end{figure}
\noindent ratus, a rotating sample stage, as well as 4-terminal dc
I-V measurements from which the film sheet resistance, R$_{\Box}$,
was obtained, have been described elsewhere \cite{Bielejec}. In
Fig. 1, we show the temperature dependence of R$_{\Box}$ for one
film section deposited on a bare glass substrate following
successive deposition steps to increase film thickness. The film
changed its behavior from insulating to superconducting when
R$_{\Box}$ at 20 K was reduced to below 10 k$\Omega$/$\Box$ with
increasing thickness. Film $\#$10 in Fig. 1, which was
superconducting with a T$_{c}$ $\sim$ 6 K, had a critical field
H$_{c}$ above the 10-T field our magnet could reach at 4.2 K. Thus
the H$_{c}$ is not far below the spin-paramagnetic limit
\cite{Fulde}, which we estimated \cite{Bielejec} to be
$\sqrt{2}\Delta$/g$\mu_{B}$ $\approx$ 11.2 T, where g $\approx$ 2
is the Land$\acute{e}$ g-factor, $\mu_{B}$ is the Bohr magneton,
and $\Delta$ = 0.92 mV is the superconducting gap for Film $\#$10.
Early studies \cite{Lazarev} estimated that the critical field was
18 $\sim$ 20 T in quench-condensed Be films of T$_{c}$ = 8 $\sim$
10 K, suggesting that these films were highly disordered with a
very short penetration depth. The data in Fig. 1 do show the
quasireentrant behavior, such as in Film $\#$7, which is typically
seen in granular films. However, this quasireentrance is seen in a
range of R$_{N}$ that is much narrower than in the case of typical
granular films \cite{White,Orr}. In addition, the T$_{c}$ of these
Be films appears to increase significantly with increasing film
thickness, which is typically seen in uniform films. Thus these Be
films show certain properties of both uniform and granular films.

In Fig. 2(a), we show the MC measured in H$_{\perp}$ at a
\begin{figure}
\centerline{\epsfig{file=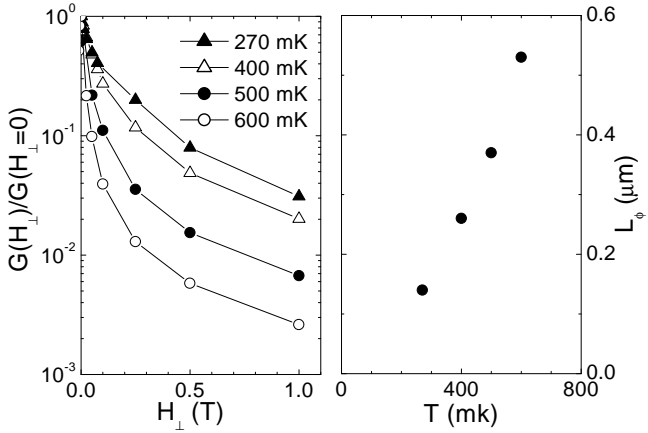,width=8.5cm}} \caption{(a):
Normalized MC measured in H$_{\perp}$ on a film similar to Film
$\#$7, in the quasireentrant regime where R$_{\Box}$ increases
with decreasing temperature. (b): The suppression of L$_{\phi}$
with decreasing temperature, obtained from the data in (a).}
\label{Figure 3}
\end{figure}

\noindent number of temperatures. In the low-field regime below 1
T, the MC is $\em{negative}$ and varies as much as seven orders of
magnitude. This can not be due to weak-localization
\cite{Bergmann}, which should lead to a $\em{positive}$ and
relatively small MC in weak spin-orbit materials such as Be. In
H$_{\parallel}$, film conductance was found to be insensitive to
the field below 1 T, as shown in Fig. 2(b). Such highly
anisotropic behavior indicates that the MC in H$_{\perp}$ is an
orbital effect. We believe that this negative MC in the low
H$_{\perp}$ regime arises as the superconducting phase coherence
is suppressed when H$_{\perp}$ exceeds the crossover value
H$^{\ast}$ that produces one flux quanta, $\Phi_{0}$ = h/2e, in a
coherent area, or when L$_{\phi}^{2}$H$_{\perp}$ $\sim$
$\Phi_{0}$. Determining this crossover field H$^{\ast}$ when the
conductance drops from its zero-field value, thus provides a
measurement of L$_{\phi}$. A few years ago, Barber and Dynes
\cite{Barber} made this argument to calculate L$_{\phi}$ in the
descending resistance tail of superconducting granular Pb films,
showing that for a superconducting film L$_{\phi}$ increases with
decreasing temperature. In our non-superconducting films $\#$7 and
$\#$8, the data plotted on a logarithmic field scale in Fig. 2(b)
show that, at 100 mK, H$^{\ast}$ was near 0.002 T for Film $\#$8
and 0.1 T for Film $\#$7, as indicate by the arrows in Fig. 2(b).
This translates into a coherence length, L$_{\phi}$, at 100 mK of
about 1.0 $\mu$m for Film $\#$8 and 0.14 $\mu$m for Film $\#$7. We
therefore see a growing L$_{\phi}$ as the films approach the
superconducting state with increasing thickness. We note that the
drop in conductance at high H$_{\parallel}$, seen in Fig. 2(b), is
likely due to the suppression of the amplitude of the
superconducting order parameter as the H$_{\parallel}$ approaches
the spin-paramagnetic limit.

Not only did we observe L$_{\phi}$ to vary with film thickness,
but it varied with temperature as well. In the temperature range
in which R$_{\Box}$ decreases with decreasing temperature, we
observed that the MC peak was sharper at lower temperatures,
indicating an increasing L$_{\phi}$ with decreasing temperature.
Such behavior is identical to that
\begin{figure}
\centerline{\epsfig{file=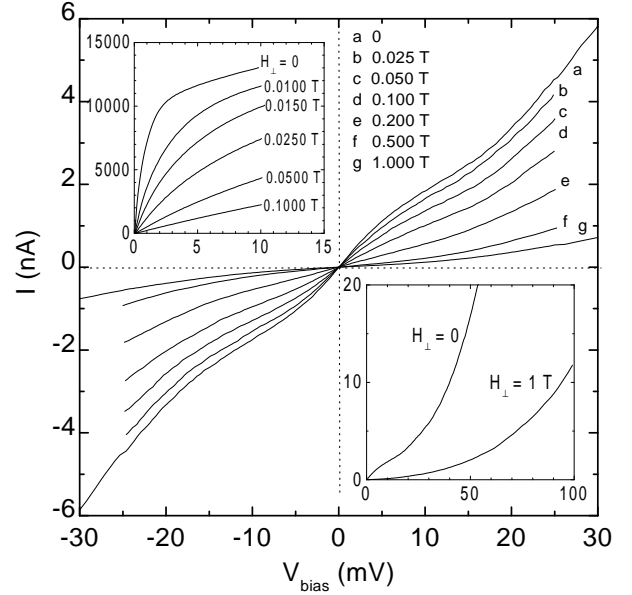,width=8.5cm}} \caption{Main
figure shows the I-V curves plotted on a low-bias scale which were
measured on insulating Film $\#$6 at 1.05 K. Inset in the
low-right corner: I-V curves on a large bias scale showing the
insulating behavior of Film $\#$6 regardless the strength of
H$_{\perp}$. Inset in the top-left corner: I-V curves measured
near the resistance minimum at 2 K of the quasireentrant Film
$\#$7, showing the suppression of much stronger supercurrent-type
behavior with increasing H$_{\perp}$. Labels on the graphs
indicate H$_{\perp}$ values in tesla.} \label{Figure 4}
\end{figure}
\noindent observed by Barber and Dynes \cite{Barber}. It is due to
the suppression of thermal fluctuations with lowering temperature.
However, we have also observed, for the first time, that the MC
peak is broader at lower temperatures in the quasireentrant regime
where R$_{\Box}$ increases with decreasing temperature, as we show
in Fig. 3 (a). Such behavior is seen in all quasireentrant films
similar to Film $\#$ 7. Using the crossover field values,
H$^{\ast}$, obtained from the data in Fig. 3 (a), we find a
reduction of L$_{\phi}$ with decreasing temperature in this
temperature range, as shown in Fig. 3 (b). We believe that this
observation demonstrates the suppression of the superconducting
phase coherence as localization effects are enhanced at lower
temperatures.

The above proposal that the MC probes the superconducting phase
coherence is further supported by the nonlinear I-V curves we have
measured near the S-I transition. Insulating and nearly
superconducting films near the S-I transition each has a distinct
type of I-V curve \cite{Orr}. In the low bias regime, the I-V
curves of nearly superconducting films show the supercurrent-type
behavior: the I-V curves have a downward curvature; while the I-V
curves of insulating films show the Coulomb-blockade-like
behavior: the I-V curves have an upward curvature. The
supercurrent-type behavior indicates the existence of a small
supercurrent associated with local superconducting regions. There
have been observations that the I-V curves evolve from the
Coulomb-blockade-type to the supercurrent-type as the films cross
the transition from the insulating side \cite{Orr}. We have seen
the same type of behavior in our films. We have also observed
that, in nearly superconducting films, the I-V curves changed from
the supercurrent-type to the Coulomb-blockade-type as the
conductance of the films is suppressed by a weak H$_{\perp}$, as
shown in the inset in the top-left corner of Fig. 4. This is
therefore additional evidence that the application of H$_{\perp}$
suppresses the superconducting fluctuations.

This negative MC and supercurrent-type I-V persisted even in much
more insulating films such as Film $\#$6, which did not show any
quasireentrant behavior. In this case, the supercurrent-type I-V
could only be observed in a narrow temperature range between 0.8
$\sim$ 1.2 K. In the main part of Fig. 4, we plot the I-V curves
measured at 1.05 K on Film $\#$6, for a number of perpendicular
field values. Although the effect was much weaker in Film $\#$6
than in less insulating films $\#$7 and $\#$8, we can see clearly
that, below a bias voltage of 15 mV, the curvature of the I-V
curves changes from downward to upward with a increasing
perpendicular field. However, as shown on a higher bias scale in
the inset in the low-right corner of Fig. 4, the I-V curves always
show an upward curvature regardless of the magnetic field,
indicating the insulating nature of Film $\#$6. Thus although
Films $\#$6 was very insulating, there still existed a finite
L$_{\phi}$, which resulted in an observable supercurrent-type I-V
in zero-field at temperatures not so low that the effect is
completely suppressed by localization.

In conclusion, we have directly observed for the first time the
finite superconducting phase coherence length L$_{\phi}$ on the
insulating side of the S-I transition. Our quench-condensed Be
films show both the quasireentrant behavior of granular films and
the varying T$_{c}$ usually seen in uniformly disordered films.
Scanning force microscopy studies have shown that the length scale
of disorder is much shorter in these Be films than that of
apparently granular films. The MC is negative, large, and highly
anisotropic in the direction of the field. Our results demonstrate
that this MC gives a direct probe of the length scale associated
with the S-I transition. In nearly insulating films, L$_{\phi}$ is
observed to $\em{decrease}$ with $\em{decreasing}$ temperature,
highlighting the competition between localization and
superconductivity. With increasing film thickness, we expect
L$_{\phi}$ to grow and to diverge as the films eventually develop
a global superconducting phase with zero resistance.

We gratefully acknowledge numerous invaluable discussions with S.
Teitel, Y. Shapir, Y. Gao, and P. Adams. We thank S. Zorba and Y.
Gao who performed scanning force microscopy studies of our
quench-condensed Be films.

%
%
%

%

%
\end{multicols}
\end{document}